\documentstyle[aps,pra]{revtex}
\begin{document}

\title{General theory and examples of the inverse Frobenius-Perron
problem}

\author{D. Pingel and P. Schmelcher}
\address{Theoretische Chemie,  Physikalisch-Chemisches Institut, Universit\"at
Heidelberg, \\
Im Neuenheimer Feld 253, D-69120 Heidelberg, Germany}

\author{F.K. Diakonos}
\address{Department of Physics, University of Athens, GR-15771 Athens, Greece}

\date{\today}
\maketitle

\begin{abstract}
The general solution of the inverse Frobenius-Perron problem considering the 
construction of a fully chaotic dynamical system with given invariant density 
is obtained within the class of one-dimensional unimodal maps. Some 
interesting connections between this solution and the approach via conjugation transformations
are eluminated. The developed method is applied to obtain a wide
class of maps having as invariant density the two-parametric beta-probability 
density function. Varying the parameters of the density a rich variety of
dynamics is observed. Observables like autocorrelation functions, power spectra
and Lyapunov exponents are calculated for representatives of this family of 
maps and some theoretical predictions concerning the decay of correlations are
tested.
\end{abstract}

\pacs{5.45+b}

\section{Introduction}

One-dimensional iterative maps represent a very useful tool to
understand the physics of complex nonlinear systems. In particular they
can be used to investigate the different routes of nonlinear
dynamical systems from regular to chaotic behaviour  \cite{Fei,Cru,Manp}
to simulate physical systems showing anomalous diffusion \cite{Geis1} or
to analyse the underlying dynamics in time series with $\frac{1}{f}$ noise 
in their power spectrum \cite{Hirs,Geis2}. In many cases and in particular from an
experimental point of view the only information
available for a physical system are its statistical properties (ergodic 
measure or time correlation function) and one is asked to design a consistent 
dynamical law. In a recent paper \cite{DiS} we examined the possibility to
construct an one dimensional discrete dynamical system with a given invariant 
density. This is the so called inverse Frobenius-Perron problem (IFPP) 
\cite{Bara,Koga,Ghik,Gros,Pala,Hunt}.

It could be shown that using certain symmetry requirements the IFPP possesses
a unique solution. We applied our method to calculate numerically the
symmetric unimodal maps with the invariant density given by the symmetric
beta distribution. However the solution presented in \cite{DiS} is a partial 
and very specialized one.
Depending on the chosen symmetry constraints we can get an infinite number
of such solutions. In the present paper we go one step further and present
the general solution to the IFPP within the class of unimodal maps. 
A central role is hereby played by the function $h_f$ which parametrizes
the space of the solutions to the IFPP.
This solution enables us to explore in depth the connection between the dynamics 
and statistics in one-dimensional systems. Our investigations concentrate on 
the wide family of the beta distributions. The IFPP is solved for the 
asymmetric case when the two exponents in the expression for the density are 
unequal. From the general solution we obtain two special solutions using two 
different forms for the function $h_f$. We then calculate the time-correlation
functions and the power spectra of the 
corresponding dynamical systems. The question of the dependence of the statistical characteristics 
(singularities of the density, long time behaviour of the autocorrelation 
function) of these systems on their dynamical characteristics (critical 
points) is addressed. Next we compare for a given density the dependence of the 
statistical properties on the form of the function $h_f$. A large
variety of the statistical and dynamical behaviour is obtained. Some theoretical 
predictions concerning the asymptotic decay of the correlations or 
the value of the Lyapunov exponents \cite{Hor,Gross,Mori,Schust} in fully chaotic 
dynamical systems can be tested within the class of our examples.

The paper is organized as follows: In section II we present the general 
solution
to the IFPP within the class of unimodal maps. In section III we derive from the
general solution of section II a wide family of unimodal maps having as 
invariant density the beta distribution. This family of maps corresponds to
two special solutions of the IFPP: the symmetric betamap (SB) and the special
asymmetric betamap (SAB). In section IV   we discuss the main characteristics
of the maps belonging to the beta family and we study the rich dynamical 
behaviour which they provide. In section V we calculate the autocorrelation 
function, the power spectrum and the Liapunov exponent for representative 
members of the beta family. A connection between the dynamics and the asymptotic decay
of correlations is established. Some theoretical models explaining this
behaviour are tested. Finally section VI contains the conclusions 
and gives a brief outlook with respect to the inclusion of the correlation
behaviour into the inverse Frobenius-Perron problem.

\section{General solution of the IFPP for unimodal ergodic and chaotic maps}

Let us consider the variety of dynamical systems which belong to an arbitrary but fixed 
invariant density. The solution to this inverse problem is of great interest
for the numerical simulation of real physical systems as well as the understanding of the
relationship between the functional form of the map and the statistical features of
its resulting dynamics. In the present section we establish a general representation of all  
ergodic and chaotic unimodal maps with a given invariant density. This corresponds
to the general representation of the
solution of the so called inverse Frobenius-Perron problem (IFPP)
\cite{Bara,Koga,Ghik,Gros,Pala,Hunt} within this class of maps.
The starting-equation for the construction of the map is the
Frobenius-Perron equation: 

\begin{equation}
\rho(y)\,|dy|=\sum\limits_{x_i=f^{-1}(y)}\rho(x_i)|dx_i| 
\label{froperequ}
\end{equation}
where the summation runs over all preimages of $y$ (for unimodal maps 
$i=L(left),R(right)$).
For any given unimodal $f(x)$ the right or left preimage of $y$ is 
determined, once the corresponding other one is given.
The essential feature of our approach is the inversion of this expression: A 
prescribed relation between the two preimages reduces the number of 
independent differentials on the rhs of eq.(\ref{froperequ}) and allows a 
following integration.
Such a relation may be described by a function $h_f(x)$, mapping the left 
preimage onto the right one:

\begin{eqnarray}
h_f&:&[0,x_{\max}]\longrightarrow[x_{\max},1] \nonumber \\
x_R&=&h_f(x_L)\;\;\;\;\;\;\mbox{with}\;\;f(x_L)=f(x_R)
\label{hfprei}
\end{eqnarray} 
where $x_{max}$ is the position of the maximum of the unimodal map.
$h_f(x)$ is a function monotonously decreasing on the defining interval
and it is differentiable with the exception of a finite number of points. 
It obeys the equations:

\begin{eqnarray}
h^\prime_f(x)<0 &&\;\;\;\;x\in[0,x_{\max}] \nonumber \\
h_f(0)=1  &&\;\;\;\;h_f(x_{\max})=x_{\max}
\label{hfprop}
\end{eqnarray}
Substituting $h_f$ in eq.(\ref{froperequ}) yields:

\begin{equation}
\rho(y)dy=\rho(x_L)dx_L-\rho(h_f(x_L))h^\prime_f(x_L)dx_L
\label{newfp}
\end{equation}
Using the definition $\mu(x)=\int_0^x \rho(t)\,dt$ for the invariant
measure eq.(\ref{newfp}) can be integrated in $0<x<x_{\max}$ providing us 
with the left part $f_L(x)=f(x)|_{[0,x_{\max}]}$ of the map $f(x)$:

\begin{equation}
f_L(x)=\mu^{-1}[\mu(x)-\mu(h_f(x))+1]
\label{leftp}
\end{equation}
with $\mu(x)$ being invertible and normalized, i.e. $\mu(1)=1$.
The right part of the map, $f_R(x)=f(x)|_{[x_{\max},1]}$, is obtained by 
substitution of $x\rightarrow h^{-1}_f(x)$ in eq.(\ref{leftp}).
It follows that the map $f(x)$ parametrised through the function $h_f(x)$ 
and characterised by the invariant measure $\mu(x)$, is given by:

\begin{equation}
f(x)=\left\{ \begin{array}{l@{\quad;\quad}l}
	\mu^{-1}[\mu(x)-\mu(h_f(x))+1]&0 \leq x<x_{\max}\\
	\mu^{-1}[\mu(h^{-1}_f(x))-\mu(x)+1]&x_{\max}\leq x \leq 1\\
             \end{array}\right\}=1-|\mu(x)-\mu(H_f(x))|
\label{funx}
\end{equation}
where $H_f(x)$ is given by:
\begin{equation}
H_f(x)=\left\{ \begin{array}{l@{\quad;\quad}l}
        h_f(x)&0 \leq x<x_{\max}\\
        h^{-1}_f(x)&x_{\max}\leq x \leq 1\\
        \end{array}\right.
\label{genhf}
\end{equation}
With the above eq.(\ref{funx}) we have found an interesting representation
and answer to the IFPP which is the central topic of this paper:
All unimodal maps with prescribed invariant measure are given by 
(\ref{funx}), $h_f(x)$ taking on all possible functional forms obeying 
(\ref{hfprei}) and (\ref{hfprop}). This shows that a fixed invariant measure
is a relatively weak constraint to a solution of the inverse problem and there
is still a considerable freedom to model the mapping. 

Before we apply our method to analyse specific examples we mention an 
interesting connection between this approach and the approach via conjugation 
transformations which can also be applied to construct a map with an 
arbitrary given (in general nonsymmetric) invariant density. To our knowledge 
the conjugation transformation method has been applied in the literature only 
to the restricted case of the construction of a symmetric map with prescribed 
invariant density \cite{Gross,Mori}. We will show below how this approach can be 
extented to include all unimodal maps. First we recall the basic features of 
the conjugation transformation approach. More details can be found in the 
literature \cite{Gyorg,Csord}.
Using as a basis the tent map given by: $t(x)=1- \vert 2 x-1 \vert$ and 
applying the conjugation transformation $g(x)=u^{-1} \circ t \circ u(x)$ with
$u(0)=0,~u(1)=1,~u'(x)>0$ for all $x \in [0,1]$ and with the additional
property $u(x)=1-u(1-x)$ (antisymmetric with respect $x_{max}=\frac{1}{2}$) we
can span the space of all doubly symmetric maps 
(symmetric maps and symmetric invariant densities). The invariant density of
the transformed map $g(x)$ is then: $\rho_g(x)=\left|\frac{du(x)}{dx}\right|$.
To arrive in an analogous way at a nonsymmetric density while the map remains 
symmetric we have to use a generalized conjugation transformation:
\begin{equation}
f(x)= U^{-1} \circ t \circ u(x)~~~~~~;~~~~U(x)=u(x)+v(x) 
\label{genconj}
\end{equation}
For symmetric $f(x)$ the function $v(x)$ must also be symmetric: 
$v(x)=v(1-x)$.
The two constraints, i.e. that the invariant density of the resulting map 
$f(x)$ is  a prescribed function $\rho(x)$ and that $f(x)$ is symmetric, 
determine uniquely the transformation functions: $u(x),v(x)$.
Defining the symmetric $\mu_+(x)=\frac{1}{2} \int\limits_0^x (\rho(t)-\rho(1-t))~dt$ and
antisymmetric parts $\mu_-(x)=\frac{1}{2} 
\int\limits_0^x (\rho(t)+\rho(1-t))~dt$ (with $\mu(x)=\mu_+(x)+\mu_-(x)$) of the
invariant measure $\mu(x)$ we find: $u(x)=\mu_-(x)~~;~~v(x)=\mu_+(x)$.
Using the fact that: $\mu_-(x)=\frac{1}{2}(\mu(x)+1-\mu(1-x))$ we get the
following expression for the desired map $f(x)$:
\begin{equation}
f(x)=\mu^{-1}\left[1-|\mu(x)-\mu(1-x)|\right]
\label{mapcon}
\end{equation}
Comparing equations (\ref{funx}) and (\ref{mapcon}) we see that they are 
compatible provided that $h_f(x)=1-x$, which corresponds to the correct choice
of $h_f$ in (\ref{funx}) for a symmetric solution $f(x)$. 

We can now extend the conjugation transformation approach in order to 
determine also nonsymmetric maps with a given invariant density. For the
derivation of eq.(\ref{mapcon}) we have used a decomposition of the 
invariant measure in a symmetric and an antisymmetric part (with respect to 
$x_{max}$). This decomposition leads us to a symmetric solution. 
A general decomposition $\mu_+(x),\mu_-(x)$ obeying 
$\mu_+(x)+\mu_-(x)=\mu(x)$ is given in the following. In the case of unimodal maps
we introduce the following components of the measure $\mu(x)$:
\begin{eqnarray}
\mu_+(x)=\frac{1}{2} \int\limits_0^x (\rho(t)+H'_f(t) \rho(H_f(t)))~dt
\nonumber \\
\mu_-(x)=\frac{1}{2} \int\limits_0^x (\rho(t)-H'_f(t) \rho(H_f(t)))~dt
\label{gendec}
\end{eqnarray}
with $H_f(x)$ given in eq.(\ref{genhf}).
Using the generalized conjugation transformation of eq.(\ref{genconj})
with $u(x)=\mu_-(x)$ and $v(x)=\mu_+(x)$ we get the solution
of eq.(\ref{funx}) to the IFPP.  Thus conjugation transformations of the
form of eq.(\ref{genconj}) with $u(x),v(x)$ chosen as in eq.(\ref{gendec}) span the 
entire space of unimodal maps with a given invariant density $\rho(x)$.

\section{An extensive class of maps with the Beta-distribution as invariant 
density}

The method to construct an one-dimensional unimodal map with a prescribed 
invariant density, as proposed
in the previous section, applies for all possible forms of the invariant measure 
$\mu(x)$ and the invariant density $\rho(x)$, respectively. Nevertheless we want 
to consider a special class of particularly relevant
invariant densities $\rho(x)$, namely the two-parametric Beta-distribution:
\begin{equation}
\rho(x)=\frac{x^{\alpha}(1-x)^{\beta}}{B(\alpha+1,\beta+1)}~~~~~~~~~~
\alpha,\beta~>~-1
\label{betden}
\end{equation}
These densities are important due to their widespread occurence and applicability (for
example in turbulent reacting flows, diffusion in disordered media, image
classification see refs.\cite{ComFlu}) as well as their simple form.
The special case of a symmetric density $(\alpha=\beta)$ with a 
symmetric solution, i.e. a map obeying $(f(x)=f(1-x))$, which is unique, has been already 
discussed in the literature \cite{DiS}. We call this
solution the doubly symmetric Beta-map (DSB). In the present paper 
we discuss the general case of an 
asymmetric invariant density $(\alpha\not=\beta)$.

We focus on two solutions of the IFPP with this
general asymmetric invariant density, both of which are identical to the DSB in 
the case $\alpha=\beta$:
\begin{itemize}
\item
The first solution is symmetric $(f(x)=f(1-x))$. We call it the 
symmetric Beta-map (SB).
\item
The second solution obeys the following constraint:
\begin{equation}
f(x_L)=f(x_R )\;\;\;\;\mbox{for }\;\; x_L,\,x_R \;\;
\mbox{with}\;\;\rho(x_L)=\rho(x_R),
\label{SAB}
\end{equation} 
which is only possible if $\alpha \cdot \beta>0$. We call this solution the 
Special Asymmetric Beta-map (SAB). In Fig.1 we show an example of such a 
map for the case $\alpha=-0.9,\beta=-0.6$. The corresponding invariant density
is also presented and the notation regarding the imposed constraint 
of eq.(\ref{SAB}) is explained.
\end{itemize}
In the following we consider only
examples where both $\alpha$ and $\beta$ possess the same sign, i.e. the 
above criterion is fulfiled.
These two types of solutions correspond of course to two different
choices for the form of the function $h_f(x)$. For the case of the SB we have
$h_f(x)=1-x$. The second case is more delicate. It 
corresponds to an implicitly defined $h_f(x)$. 
From the condition in eq.(\ref{SAB}) and eq.(\ref{betden}) we find: 
\begin{equation}
(1-x_R) x_R^{\gamma}=(1-x_L) x_L^{\gamma}
\label{hfimp}
\end{equation}
where $\gamma=\frac{\alpha}{\beta}$. Eq.(\ref{hfimp}) defines $h_f$ and can be solved 
parametrically. Using $\lambda=\frac{x_L}{x_R}$ we get:
\begin{equation}
x_L=\lambda\frac{1-\lambda^{\gamma}}{1-\lambda^{\gamma+1}}~~~~~~~;~~~~~~~
x_R=\frac{1-\lambda^{\gamma}}{1-\lambda^{\gamma+1}}
\label{deflam}
\end{equation}
with $0 \leq \lambda \leq 1$.
Let us now apply the theoretical considerations of section II to derive the
expressions for the two types of Beta-maps refered above.
The invariant measure corresponding to the density (\ref{betden}) is:
\begin{equation}
\mu_B(x)=\frac{\int\limits_0^x dt~t^{\alpha}(1-t)^{\beta}}
{B(\alpha+1,\beta+1)}=
\frac{B(\alpha+1,\beta+1,x)}{B(\alpha+1,\beta+1)}
\label{betmeas}
\end{equation}
where $B(a,b,z)$ is the incomplete beta function \cite{GradRez}.
Using eq.(\ref{leftp}) we can determine implicitly the left branch of the 
solution of the IFPP:
\begin{equation}
B(\alpha+1,\beta+1,f_L(x))=B(\alpha+1,\beta+1,x)-B(\alpha+1,\beta+1,h_f(x))
+B(\alpha+1,\beta+1)
\label{bleft}
\end{equation}
and a similar expression results also for $f_R(x)$: 
\begin{equation}
B(\alpha+1,\beta+1,f_R(x))=B(\alpha+1,\beta+1,h_f^{-1}(x))-
B(\alpha+1,\beta+1,x)+B(\alpha+1,\beta+1)
\label{bright}
\end{equation}
To get the two special solutions refered above we have to substitute in 
eqs.(\ref{bleft},\ref{bright}) the suitable $h_f$. As we already mentioned
for the SB we choose $h_f(x)=h_f^{-1}(x)=1-x$. Taking into account this
special form of $h_f$ and the fact that the 
eqs.(\ref{bleft},\ref{bright}) are valid for $x < x_{max}$ and $x > x_{max}$,
respectively (for SB $x_{max}=\frac{1}{2}$), we can recast eqs.(\ref{bleft},
\ref{bright}) into a single equation: 
\begin{equation}
B(\alpha+1,\beta+1,y_{SB})=B(\alpha+1,\beta+1)-\vert 
B(\alpha+1,\beta+1,x)+B(\beta+1,\alpha+1,x)-B(\alpha+1,\beta+1) \vert~~~~
x \in [0,1]
\label{SBmap}
\end{equation}
with $y_{SB}=f_{L,SB}(x)$ for $0 \leq x \leq x_{max}$ and $y_{SB}=f_{R,SB}(x)$
for $x_{max} < x \leq 1$. To calculate the value of the SB  for a given x
we have to solve eq.(\ref{SBmap}) numerically using a standard root finding 
routine. It turns out that the bisection method \cite{Recip} is well suited 
for a fast and accurate determination of $y_{SB}$. In Fig.2a we present
4 characteristic members of the class of symmetric beta-maps calculated 
numerically using (\ref{SBmap}).

Let us now return to the SAB case. The calculation of the map $y_{SAB}$ is a
little more tedious. For a given $x$ we use eq.(\ref{hfimp}) to find 
numerically $x_R$ ($x_L$) if $x < x_{max}$ (if $x > x_{max}$). Note that for 
the SAB $x_{max}=\frac{\alpha}{\alpha+\beta}$. The numerical solution to 
eq.(\ref{hfimp}) can be easily obtained using the bisection method. 
Then we apply eq.(\ref{funx}) to determine $y_{SAB}$:
\begin{equation}
B(\alpha+1,\beta+1,y_{SAB})=\left\{ \begin{array}{l@{\quad;\quad}l}
	B(\alpha+1,\beta+1,x)-B(\alpha+1,\beta+1,x_R)+B(\alpha+1,\beta+1)&
        0 \leq x<x_{\max}\\
	B(\alpha+1,\beta+1,x_L)-B(\alpha+1,\beta+1,x)+B(\alpha+1,\beta+1)&
        x_{\max}\leq x \leq 1\\
        \end{array}\right.
\label{SABmap}
\end{equation}
Again we have to find the root $y_{SAB}$ of eq.(\ref{SABmap}) numerically
using the bisection method.
We emphasize that the bisection procedure in solving eq.(\ref{SABmap})
could be circumvented in the case that the inverse of the incomplete beta function
$B^{-1}(a,b,z)$ is given explicitly.
The calculated SAB maps, for the same choice of the values of $\alpha$
and $\beta$ as in Fig.2a, are presented in Fig.2b. 

For both types of maps SB and SAB a dynamical
route to the fully chaotic and ergodic state can be constructed applying the
method described in \cite{DiS}. Using the numerical solutions to 
eqs.(\ref{SBmap},\ref{SABmap}) we calculated trajectories for the 
corresponding
maps for different values of the parameters $\alpha$ and $\beta$. 
Averaging over a long trajectory of each map we reproduced with a good 
agreement the invariant densities given as the input for the construction of 
these maps. We mention that, due to computational accuracy reasons,
for $\alpha,\beta$ values close to $-1$ it is necessary to 
calculate $1-y$ instead of $y$ in that range of $x$ for which
$y(x) \approx 1$.

Finally let us introduce the concept of the complementary 
beta-map, CB. Two maps $y(x)$ and $\tilde{y}(x)$ belonging to
SB or SAB are called complementary, if their invariant densities $\rho(x)$ and
$\tilde{\rho}(x)$ are transformed into each other
by changing $x$ to $1-x$. Considering the Frobenius-Perron equation, which is the 
starting point for the construction of both SB and SAB maps, 
the following relation can be derived: 
\begin{equation}
B\left[\alpha+1,\beta+1,y(x)\right]=
B\left[\beta+1,\alpha+1,\tilde{y}(1-x)\right]
\label{betcom}
\end{equation}
Complementary maps agree with respect to certain dynamical features, e.g. the Lyapunov-exponent,
as shown below.

\section{Analytical and dynamical properties of the SB and SAB}

Many features of the dynamics generated by a discrete map do not depend on the
dynamical law as a whole, but only on its local behaviour in certain regions 
of the phase space. In particular for the case of unimodal maps the order of 
the maximum turns out to be responsible for some interesting quantities 
characterising the underlying dynamics like the singularities of the invariant
density at $x=0$ and $x=1$, the decay constant of the autocorrelation and the 
scaling parameter of the Feigenbaum bifurcation route \cite{Ott}.
The order $\nu_m$ of a map at its maximum $x_{\max}$, is defined as 
the exponent in the approximation: 
\begin{equation}
f(x)\approx 1-(x-x_{\max})^{\nu_m} \;\;\;\;\;\;\;
\mbox{for}\;\;\;\;\;x\approx x_{\max}\
\end{equation}
In an analogous way we can define the order of the map at the end points:
$x=0,1$. In the following we will determine the corresponding exponents 
$\nu_m,\nu_0,\nu_1$ for the SB and SAB.\\

\noindent {\bf{(a) The SB case}}\\

\noindent One can easily utilise the Frobenius-Perron-equation 
\begin{equation}
\frac{dy}{dx} =\frac{x^\alpha(1-x)^\beta+x^\beta(1-x)^\alpha}
{y^\alpha(1-y)^\beta}
\end{equation}
where $y=f(x)$ is the SB-map. For 
$x\rightarrow x_{\max}=\frac{1}{2}$ the nominator on the rhs is finite which yields
$\frac{dy}{dx} \propto\frac{1}{y^\alpha(1-y)^\beta}$
Assuming $\nu_{m,SB}$ to be the order of the map in the maximum, i.e.
$(1-y)\propto \left(x-\frac{1}{2}\right)^{\nu_{m,SB}}$ one gets
\begin{equation}
\nu_{m,SB}=\frac{1}{\beta+1}
\label{orSB}
\end{equation}
Expanding eq.(\ref{SBmap}) around the points $x=0$ and $x=1$ and taking into
account that $y_{SB} \to 0$ for $x \to 0$ or $x \to 1$ we find the exponents
$\nu_{0,SB}$ and $\nu_{1,SB}$ to be:
\begin{equation}
\nu_{0,SB}=\nu_{1,SB}=\frac{\min(\alpha,\beta)+1}{\alpha+1}
\label{SBor01}
\end{equation}

\noindent {\bf{(b) The SAB case}} \\

\noindent Determining the order of the SAB at its maximum 
$x=\frac{\alpha}{\alpha+\beta}$ turns out to be slightly more complicated
compared to the case of the SB due to the implicit form of $h_f$. 
The derivative $\frac{dy}{dx}$ of the SAB
$y=f(x)$ is given by the Frobenius-Perron equation. In terms of the parametric
solution $x_R(\lambda)$, $x_L(\lambda)$ of eq.(\ref{deflam}) we get:
\begin{equation}
\frac{dy(x)}{dx}=\left\{
\begin{array}{c@{\quad,\quad}c}
{\frac{dy(\lambda)}{d\lambda}}/{\frac{dx_R}{d\lambda}} & 0<x<x_{max}\\
{\frac{dy(\lambda)}{d\lambda}}/{\frac{dx_L}{d\lambda}} & x_{max}<x<1\\
\end{array}\right\}
\end{equation}
The limit $(x-x_{max})\rightarrow 0$ corresponds to the limit of the above
equation for $\lambda\rightarrow 1$. 
Higher derivatives $\frac{d^ny}{dx^n}$ are obtained by differentiating 
$\frac{d^{n-1}y}{dx^{n-1}}$ with respect to $\lambda$ and subsequent dividing 
by $\frac{dx_L}{d\lambda}$ or $\frac{dx_R}{d\lambda}$, respectively.
The leading behaviour of the left or right derivative 
$\frac{d^ny}{dx^n} \vert_{L,R}$ for $y \to 1$ (which corresponds to
$(x-x_{max})\rightarrow 0$) is given by:
\begin{equation}
\frac{d^ny}{dx^n} \vert_{L,R} \sim \frac{\frac{d^ny}{d\lambda^n}}
{(\frac{dx_{L,R}}{d\lambda})^n}
\label{leadbeh}
\end{equation}
A straightforward but tedious calculations shows that
$\frac{d^ny}{dx^n}\propto\frac{1}{(1-y)^{n\beta+(n-1)}}$
The exponent of $(1-y)$ vanishes for a value 
\begin{equation}
n=\nu_{m,SAB}=\frac{1}{\beta+1}
\label{orSAB}
\end{equation}
which gives the order $\nu_{m,SAB}$ of the SAB at the maximum.
Expanding eq.(\ref{SABmap}) around the endpoints $x=0,1$ we get:
\begin{equation}
\nu_{0,SAB}=\min(1,\frac{\alpha (1+\beta)}{\beta (1+\alpha)})~~~~~;~~~~~
\nu_{1,SAB}=\min(\frac{\beta}{\alpha},\frac{1+\beta}{1+\alpha})
\label{SABor01}
\end{equation}

Acoording to the behaviour of the SB and the SAB near the characteristic points 
$0,1,x_{max}$ we can classify the forms of maps which we obtain for different
values of $\alpha$ and $\beta$ as follows:
\begin{itemize}
\item The region $x \approx x_{max}$:
The order of the map at $x_{max}$, given by $\nu_m=\frac{1}{1+\beta}$, 
is a common quantity for all beta-maps. Maps with $\beta<0$ are differentiable
and therefore flat at $x_{max}$. With $\beta$ approaching $-1$, the 
maps become increasingly flatter. $\beta>0$ results in a cusp-like shape of the map 
near $x_{max}$. 

\item The zero points $0$ and $1$:

As far as the form of the maps near the points $1$ and $0$ is concerned, four 
limiting cases with certain values for $\alpha$ and $\beta$ can distinguished.

\begin{itemize}
\item[-] $\alpha\approx 0$, $\beta \approx 1$:

The maps belonging to both SAB and SB show divergent derivatives
$f^\prime(0)\longrightarrow \infty,~ f^\prime(1) 
\longrightarrow -\infty$. The map itself has a convex shape. As the maximum of the SAB is in 
$[0,\frac{1}{2}]$, the right branch of the map, containing the unstable fixed 
point, is more shallow than that of the corresponding SB map.

\item[-] $\alpha\approx 1$, $\beta \approx 0$:

At $x=0$, both the SB as well as the SAB map develop a slope 
$ f^\prime(0) \longrightarrow 1 $
which enables them to generate intermittent dynamics. Since the SB are 
symmetric we have $ f^\prime(1) \longrightarrow 1$.
The derivatives of the SAB maps however show divergent behaviour,
$ f^\prime(1) \longrightarrow -\infty $

\end{itemize}

The SB and SAB maps are identical to the DSB for $\alpha=\beta$. The derivatives of 
the maps $ f^\prime(0)$ and $f^\prime(1)$ take on finite values.
\begin{itemize}
 
\item[-] $\alpha=\beta\approx 0$:
The limit $\alpha=\beta\longrightarrow 0$ of the DSB is the tent map
$ f^\prime(0) \longrightarrow 2,f^\prime(1) \longrightarrow -2 $.

\item[-] $\alpha=\beta\approx 1$:

The values of $f^\prime(0) = f^\prime(1)$ are still finite, but increase 
unbounded in the limit $\alpha=\beta\longrightarrow 1$. The map is extremely 
flat about the maximum and takes on values very close to $1$ in nearly
the whole interval $[0,1]$. 

\end{itemize}

\end{itemize}

The dynamics generated by both the SAB and the SB result in singularities of 
the invariant density at $x=0,1$ for $-1~<~\alpha,\beta~<~0$.  Although the 
strengths of these singularities can be directly read from eq.(\ref{betden})
it is of interest to explore how they are determined through the order of the 
map at the maximum and at the endpoints $0,1$. The strength of the singularity
at $x=1$ is determined by the order of the maximum only. 
As the order of the map for both SB and SAB at its maximum is given by 
$\nu_m=\frac{1}{\beta+1}$, the expansion of the map in this region reads
\begin{equation} 
1-y \approx \left|1-x_{max}\right|^{\nu_m}=
\left|1-x_{max}\right|^{\frac{1}{\beta+1}}
\end{equation}
Near $x_{max}$ the invariant measure is a smooth function. Using eq.(\ref{froperequ})
we can therefore write $\rho(y)\:|dy|\propto \:|dx|$.
The dependence of the differential $|dy|$ on $x$ is given by
$|dy|\approx\frac{1}{\beta+1}\left(x-\frac{1}{2}\right)^{\frac{1}{\beta+1}-1}
\:|dx|$. If we now make the ansatz $\rho(y) \;\propto\;(1-y)^{\tilde{\beta}}$
for the invariant density at $x\approx 1$, the Frobenius-Perron equation in 
orders of $x$ reads 
\begin{equation}
\left(x-x_{\max}\right)^{\tilde{\beta}\frac{1}{\beta+1}}
\left(x-x_{\max}\right)^{\frac{1}{\beta+1}-1}|dx|
\;\;\;\propto\;\;\; |dx|\nonumber\\
\Longrightarrow \;\;\;\tilde{\beta}\frac{1}{\beta+1}+\frac{1}{\beta+1}-1=0
\nonumber\\
\Longrightarrow \;\;\;\tilde{\beta}=\beta
\end{equation}
We therefore conclude that the strength of the singularity of the invariant 
density at $x=1$ for both types of maps (SB and SAB) is given implicitly
by the order of the dynamic law at its maximum.

The determination of the singularity at $x=0$ is more intricate. We will study in 
detail the case of the SB maps. The SAB maps can be treated in an analogous 
way. For the SB family the Frobenius-Perron equation can be written in the 
form:
\begin{equation}
\rho(y_{SB})\:|dy_{SB}|=\left[\rho(x)+\rho(1-x)\right]\:|dx|,
\end{equation}
Using eq.(\ref{SBor01}) the SB at $x\approx 0$ is approximately given by
$y_{SB} \approx x^{\frac{\min(\alpha,\beta)+1}{\alpha+1}}$.
Let us make the ansatz $\rho(y_{SB}) \;\propto\;y_{SB}^{\tilde{\alpha}}$ 
for the singularity of the invariant density at $y_{SB}=0$.
For $x\approx 0$  $\rho(x)$ and $\rho(1-x)$ possess the singularities  
$\tilde{\alpha}$ and $\beta$, respectively. Therefore the Frobenius-Perron 
equation reads
\begin{equation}
x^{\tilde{\alpha}\frac{\min(\alpha,\beta)+1}{\alpha+1}}
x^{\frac{\min(\alpha,\beta)+1}{\alpha+1}-1}|dx|\propto x^{\min(\tilde{\alpha},
\beta)}|dx|
\end{equation}
\begin{equation}
\Longrightarrow \;\;\;\tilde{\alpha}\;\frac{\min(\alpha,\beta)+1}{\alpha+1}+
\frac{\min(\alpha,\beta)+1}{\alpha+1}-1=\min(\tilde{\alpha},\beta)
\label{SBsi0}
\end{equation}
Two cases have to be considered:
\begin{itemize}
\item
The first possibility is $\min(\alpha,\beta)=\beta$. In this case eq.(\ref{SBsi0}) yields
$\tilde{\alpha}(\beta+1)+(\beta-\alpha)=(\alpha+1)\min(\tilde{\alpha},\beta)$ and therefore
\begin{equation}
\min(\tilde{\alpha},\beta) = \frac{\beta(\tilde{\alpha}+1)+(\tilde{\alpha}-\alpha)}{\alpha+1}
\min(\tilde{\alpha},\beta)
\end{equation}
with the solution $\alpha=\tilde{\alpha}$. The singularity at $x=0$ is given by 
the parameter $\alpha$ as expected.

\item

The case $\alpha<\beta$ demands slightly more attention. The order of the map 
in $x=0$ and $x=1$ is one, according to eq.(\ref{SBor01}). This makes the 
dynamics a candidate for intermittency, which is characterised by an expansion
of the map near a marginally unstable fixed point of the form 
\begin{equation}
y(x)\approx x+\epsilon x^z\;\;\;\;\;\;x\rightarrow 0
\label{intmap}
\end{equation} 
The order of the next-to leading term, $z$, determines essentially the 
statistical properties of the dynamical law near $x=0$.
In ref.\cite{Hor} it was shown, that a symmetric map, with a maximum of order 
$\nu_m$ and of the form (\ref{intmap}) at $x=0$, generates an invariant 
density, which has a singularity of the order
\begin{equation} 
\tilde{\alpha}=\frac{1}{\nu_m}-z
\label{sing0}
\end{equation} 
at $x=0$. Using the implicit representation of the SB map:
\begin{equation}
B(\alpha+1,\beta+1,y_{SB})=B(\alpha+1,\beta+1,x)+B(\beta+1,\alpha+1,x),
\;\;\;\;\;\;x<\frac{1}{2}
\end{equation} 
and expanding the incomplete Beta-functions around $x=0$ with the help of the 
formula:
\begin{equation}
B(\alpha+1,\beta+1,t)\approx \frac{t^{\alpha+1}}{\alpha+1}-
\beta\frac{t^{\alpha+2}}{\alpha+2}
\end{equation} 
we find:
\begin{equation}
\frac{y_{SB}^{\alpha+1}}{\alpha+1}\approx\frac{x^{\alpha+1}}{\alpha+1}+
\frac{x^{\beta+1}}{\beta+1}
\end{equation} 
It follows that:
\begin{equation}
y_{SB} \approx x \left(1+\frac{\alpha+1}{\beta+1}
x^{\beta-\alpha}\right)^{\frac{1}{\alpha+1}} 
\approx x+\frac{1}{\beta+1}x^{\beta-\alpha+1}
\end{equation} 
Therefore the term next to the leading one is of order $z=\beta-\alpha+1$. 
Using eq.(\ref{sing0}) and $\nu_m=\frac{1}{\beta+1}$ we find 
$\tilde{\alpha}=\beta+1-(\beta-\alpha+1)=\alpha$,
i.e. the correct strength of the singularity at $x=0$ is reproduced.
\end{itemize} 

At the end of this section let us discuss the different kinds of dynamics 
which are generated by the different types of maps in the two classes
SAB and SB. Depending on the values of $\alpha$ and $\beta$, there are two 
eyecatching types of dynamical behaviour: 
\begin{itemize}
\item
$0>\alpha>\beta$

The trajectories of both SB and SAB maps are characterised by an oscillatory 
motion around the unstable fixed point in the right interval of monotony of the
map. The orbit leaves the region close to the fixed point, where the linear 
approximation of the map is valid, exponentially fast. This phase of the dynamics
is highly correlated.
The time which a corresponding trajectory spends in the neighbourhood of the
unstable fixed point is for SAB maps typically larger than for SB maps.
The reason for this is the fact that the derivations of the SAB maps are usually
in a larger region closer to $-1$ than those of the SB maps.
After leaving this regime we encounter an oscillatory motion of the orbit about the maximum. In this case the left 
and the right branch of the map are visited alternately. As the parts of the 
map visited are nearly symmetric with respect to the 
axis $y(x)=x$, a nearly stable orbit with period two occurs. This peculiar 
feature of the map is more common for the SB maps. The two singularities of 
the invariant density are build up by these oscillations. As the centre of the 
oscillations is shifted to the right of $\frac{1}{2}$, the singularity at $x=1$ is stronger 
than the one at $x=0$.

\item
$\alpha<\beta<0$

The dynamics developed by both the SB and the SAB maps is dominated by the 
intermittent behaviour near the marginally unstable fixed point at 
$x=0$. The trajectories show an interplay of chaotic motion and laminar phases
in the region of $x=0$. The comparison of the maps belonging to the two 
classes shows nicely the detailed balance of the system generating the 
same invariant density: For a SAB the exponent $z$ in the expansion 
$y(\epsilon)\approx x+\epsilon^z$, $\epsilon\ll 1$ is
much closer to $1$ than for the corresponding SB map and the region of the map
where this expansion holds is larger. Therefore the orbit tents to be captured for
a longer time in this region. However, due to the diverging derivative of the 
SAB  map at $x=1$, the laminar region is visited less often, which partially 
cancels the above effect.
\end{itemize}

\section{Lyapunov exponents, autocorrelations and power spectra for SB and 
SAB}

In this section we derive some dynamical and statistical properties of
characteristic members of the SB and SAB classes of maps. First we consider 
the Lyapunov-exponents which are a measure for the degree of chaoticity.
As the derivatives of both classes SB and SAB are 
obtained from the Frobenius-Perron equation, the Lyapunov-exponents 
$\Lambda_{\alpha,\beta}^{SB}$ and $\Lambda_{\alpha,\beta}^{SAB}$, respectively
can be calculated using the definition
$\Lambda=\int\limits_0^1\rho(x) \ln\left|\frac{dy(x)}{dx}\right|\;dx$.
The implicit form of $y(x)$ prevents an analytical treatment. 
Therefore we have calculated $\Lambda_{\alpha,\beta}^{SB}$ and 
$\Lambda_{\alpha,\beta}^{SAB}$  numerically 
for different values of $-1<\alpha,\beta<0$, using the equations 
(\ref{SBmap}) and (\ref{SABmap}), respectively.
Both Lyapunov-exponents show similar features as functions of $\alpha$ and 
$\beta$. Most striking is the symmetry
\begin{equation}
\Lambda^{SB}_{\alpha,\beta}=\Lambda^{SB}_{\beta,\alpha}~~~~~~~~~
\Lambda^{SAB}_{\alpha,\beta}=\Lambda^{SAB}_{\beta,\alpha}
\end{equation}
i.e. the Lyapunov-exponent stays the same when switching from Beta-maps to 
the complementary ones, given in eq.(\ref{betcom}).
This symmetry is even more astonishing, since it connects two very different
kinds of dynamics with the same Lyapunov-exponent, i.e. with the same degree 
of chaoticity. The value of the Lyapunov exponent is maximal ($ln2$) for the 
doubly symmetric case $\alpha=\beta$ and decreases as  
$\vert \alpha-\beta \vert$ increases. This can be at best seen in Fig.3 where
we show the Lyapunov exponent $\Lambda^{SAB}_{\alpha,\beta}$ as a function
of the parameters $\alpha,\beta$. The decrease in the degree of chaoticity 
for the case $\alpha<\beta$ is caused by the frequent occurrence of almost regular intermittent 
intervals while in the case $\alpha>\beta$ it is caused by the dominating hopping of the 
orbit around the unstable fixed point of the map. 

Let us now turn to the calculation of the autocorrelation function $C(n)$
for the SB and SAB members. Again two types of maps in the SB and SAB classes have to be 
distinguished:
\begin{itemize}
\item
$\alpha>\beta$

The correlation function is oscillating with period $2$ and decays 
exponentially fast. This is a consequence of the oscillatory motion of the 
orbit about the unstable fixed point. The decay of the envelope is slower 
for maps with $|\beta-\alpha|$ large, since the corresponding map has a more 
shallow slope of the right interval of monotony, which allows long lasting 
oscillations. The mean decay time of the SB maps (see Fig.4a) is much shorter 
compared to those of the SAB maps (see Fig.4b). This is caused by the fact that the 
symmetry of the map restricts the length of the highly correlated, oscillating
phases of the orbit. 

\item
$\alpha<\beta$

The intermittent dynamics results in a correlation function, which decays 
algebraically in the long time regime according to 
$C(n)\propto n^{-\eta}$. For weak intermittent dynamics the decay exponent
$\eta$ can be estimated analytically. For this aim we use the map:
\begin{equation}
y(\epsilon)=\epsilon+a\:\epsilon^z+\ldots\;\;\;\;\;\;\;\;\;\;\;
0<\epsilon\ll 1
\label{inma}
\end{equation}
Assuming that the invariant density $\rho(x)$ near the centre of 
intermittency is given by
$\rho(\epsilon)\propto\epsilon^\alpha\;\;\;\;0<\epsilon\ll 1,-1<\alpha,1<z$
and following \cite{Hor} we derive the following scaling relation describing 
the autocorrelation function in the limit of distinctive intermittent 
behaviour, i.e. $0<z-1 \ll 1$ and for $n\rightarrow \infty$:
\begin{equation}
C(n)\approx K \left(  1+\frac{n}{n_0}\right)^{-\eta}
\end{equation}
with
\begin{equation}
\eta=\frac{1+\alpha}{z-1}\;\;\;\;\;>0~~~~~and~~~~~\frac{1}{n_0}=(z-1)\:a\:(\bar{x}+1)^{z-1}
\end{equation}
where $K$ is a constant characteristic for the system. 
This behaviour takes account of the long and highly correlated laminar 
sections of the orbit. The characteristic exponent $\eta$ decreases with 
increasing $|\beta-\alpha|$ and thus corresponds to 
a slower decay of the correlation. For shorter times a transient behaviour 
leading into the algebraic decay can be seen. 
The correlation of systems generated by SB maps (Fig.5a) decay faster compared
to those SAB maps with the same values for $\alpha$ and $\beta$ (Fig.5b). 
The knowledge of the order of the SB maps in the maximum point and the zeros 
at $x=0$ and $x=1$ allows a comparison of the 
characteristic exponent $\eta_{numer}$ obtained numerically with the value 
$\eta_{approx}$ obtained by the continuous time approximation in \cite{Hor}.
It shows a fair agreement that becomes better for maps which possess
a more pronounced intermittency. In this case the values of $\alpha$ and $\beta$ differ 
most. 
\end{itemize}

\vspace*{2.0cm}

\begin{table}
	\begin{center}
\begin{tabular}{|c|c|c|c|c|}\hline
$\alpha$&$\beta$ & $z$ & $\eta_{approx}$&$\eta_{numer}$\\ \hline\hline
-0.8&-0.6&1.2&1.0&1.09\\  \hline
-0.6&-0.4&1.2&2.0&1.86\\  \hline
-0.6&-0.2&1.4&1.0&0.81\\  \hline
-0.8&-0.2&1.6&0.33&0.33\\ \hline
\end{tabular}
	\end{center}
\caption{Decay constants $\eta_{approx}$ and $\eta_{numer}$}
\end{table}

The corresponding power spectrum of both the SB and SAB maps, defined by
\begin{equation}
S(f)=\frac{1}{\sqrt{2 \pi}}\int\limits_{-\infty}^{\infty}C(\tau)
e^{-2\pi i f\tau}\;d\tau
\end{equation}
shows the typical power-law scaling behaviour: 

\begin{equation}
S(f) \propto f^{-\eta}
\end{equation}
for low frequencies. For higher frequencies $S(f)$ differs from this rule 
since it reflects the transient regime of the correlation function. This 
behaviour of the power spectrum in the intermittent regime is illustrated in 
Fig.6a,b for different sets of $\alpha,\beta$.

\section{Conclusions}

We have presented a general solution to the inverse Frobenius-Perron problem
for the case of unimodal one-dimensional maps. This solution allows us to
cover the entire space of fully chaotic unimodal maps with given invariant
density. It is shown that the general solution can also be obtained by
a suitably defined conjugation transformation. Using this general 
representation one can find infinitely many particular solutions. To illustrate our
method we applied it to obtain two classes of dynamical systems possessing 
the two parametric beta-probability function as the invariant density. We 
calculated the Lyapunov exponents, autocorrelation functions and power spectra
for these maps and studied the properties of the obtained solutions in a twofold manner:
First we observed for given values of the parameters of the invariant density 
the changes in the dynamical and statistical properties for
systems corresponding to different partial solutions of the Frobenius-Perron
equation. We then varied the parameters of the invariant density remaining 
within a certain class of partial solutions. From our studies we conclude
that two typical properties dominate the dynamics for unimodal beta maps.
The first property is the hopping of the chaotic trajectory around the unstable fixed point of
the map which leads to an exponential decay of correlations. The decay constant
depends on the explicit form of the special solution for given parameters
of the invariant density. The second possibility is the intermittent scenario:
the chaotic trajectory possesses extented laminar phases in the neighbourhood of
$x=0$. The correlations in this case decay with a power-law. Theoretical
investigations predict that the characteristic exponent in this power-law 
decay is determined through the singularities of the
invariant density at the center of intermittency ($x=0$) as well as the 
derivative of the map at this point. We tested these estimations and found 
a satisfactory agreement with our numerical calculations. 

Thus our approach to the inverse Frobenius-Perron problem allows us to
determine one-dimensional maps with a rich variety of dynamical behaviour and 
enables us to perform accurate calculations for physical observables even in 
the case where the ergodic limit is computationally inaccessible (intermittency).
The obtained general solution involves the function $h_f$ and one might therefore
use this degree of freedom to design dynamical laws with more prescribed quantities
than the invariant density.
One step in this direction would be to solve the inverse Frobenius-Perron 
problem with the additional constraint of a given autocorrelation function. Using the
parametrisation (\ref{funx}) of the general solution to the Frobenius-Perron
equation it should be in principle possible to design a map which is closest
to a certain given correlation behaviour. A detailed 
investigation of this problem is in progress \cite{DiPS}.

{}

\vspace*{1.0cm}

\begin{center}
{{\bf{FIGURE CAPTIONS}}}
\end{center}

\begin{figure}
\caption{The SAB map for $\alpha=-0.9,\beta=-0.6$ and the corresponding invariant density.}
\end{figure}
\begin{figure}
\caption{Characteristic members of the family of maps with the 
beta-distribution as invariant density:
(a) 4 Maps belonging to the SB class.
(b) The corresponding 4 maps of the SAB class.}
\end{figure}
\begin{figure}
\caption{The Lyapunov exponent $\Lambda_{\alpha,\beta}$ for the SAB class as a
function of $\alpha,\beta$.} 
\end{figure}
\begin{figure}
\caption{The absolute value of the autocorrelation function 
$\vert C_{\alpha,\beta} \vert$ for different values of $\alpha,\beta$ and in the regime $\alpha~>~\beta$
(exponential regime):  
(a) for SB maps. 
(b) for maps belonging to the SAB class.}
\end{figure}
\begin{figure}
\caption{The absolute value of the autocorrelation function 
$\vert C_{\alpha,\beta} \vert$ for different values of $\alpha,\beta$ and for $\alpha~<~\beta$ 
(intermittent regime):  
(a) for SB maps. 
(b) for maps belonging to the SAB class.}
\end{figure}
\begin{figure}
\caption{The square of the absolute value of the power spectrum $S(f)$ 
as a function of the frequency $f$ for different values of $\alpha,\beta$ and 
for $\alpha~<~\beta$ (intermittent regime):  
(a) for SB maps. 
(b) for maps belonging to the SAB class.}
\end{figure}

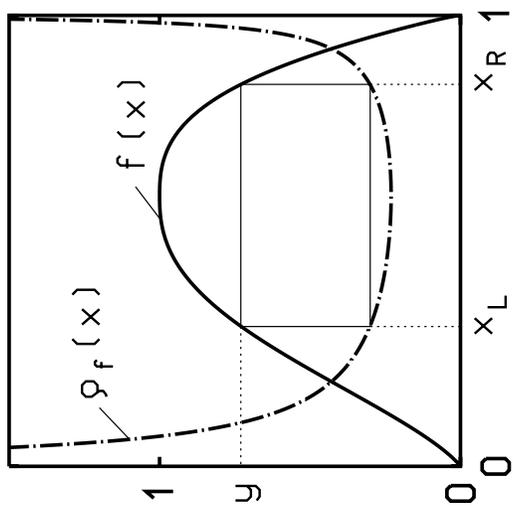
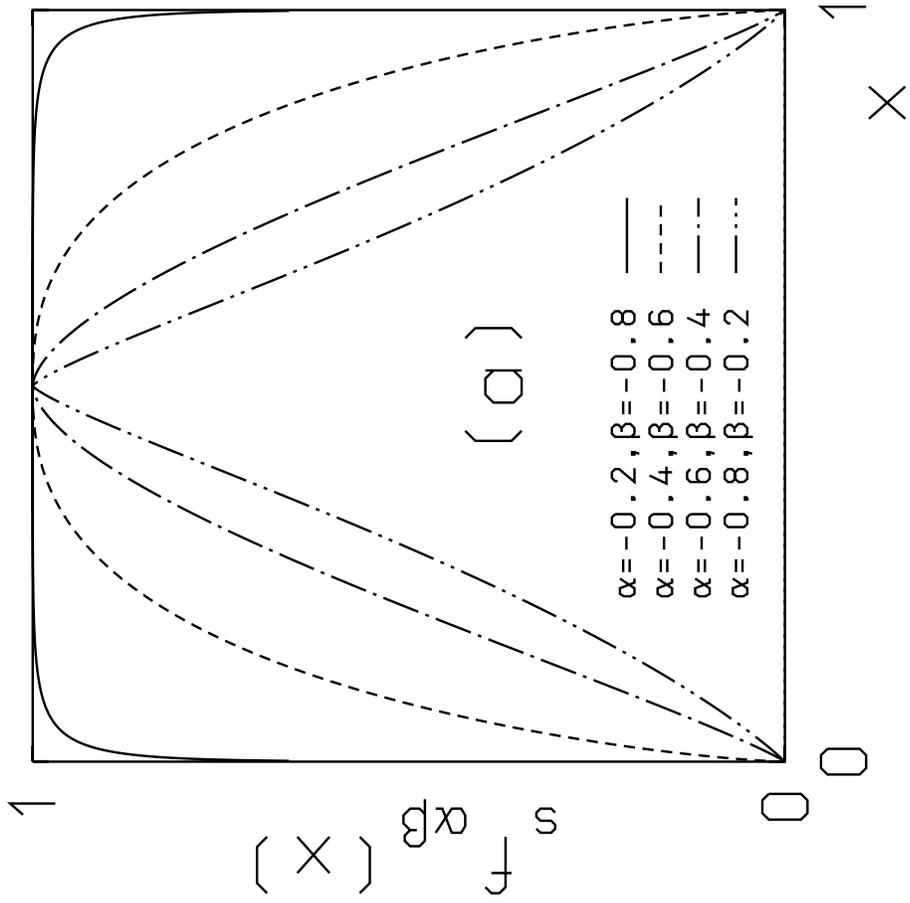
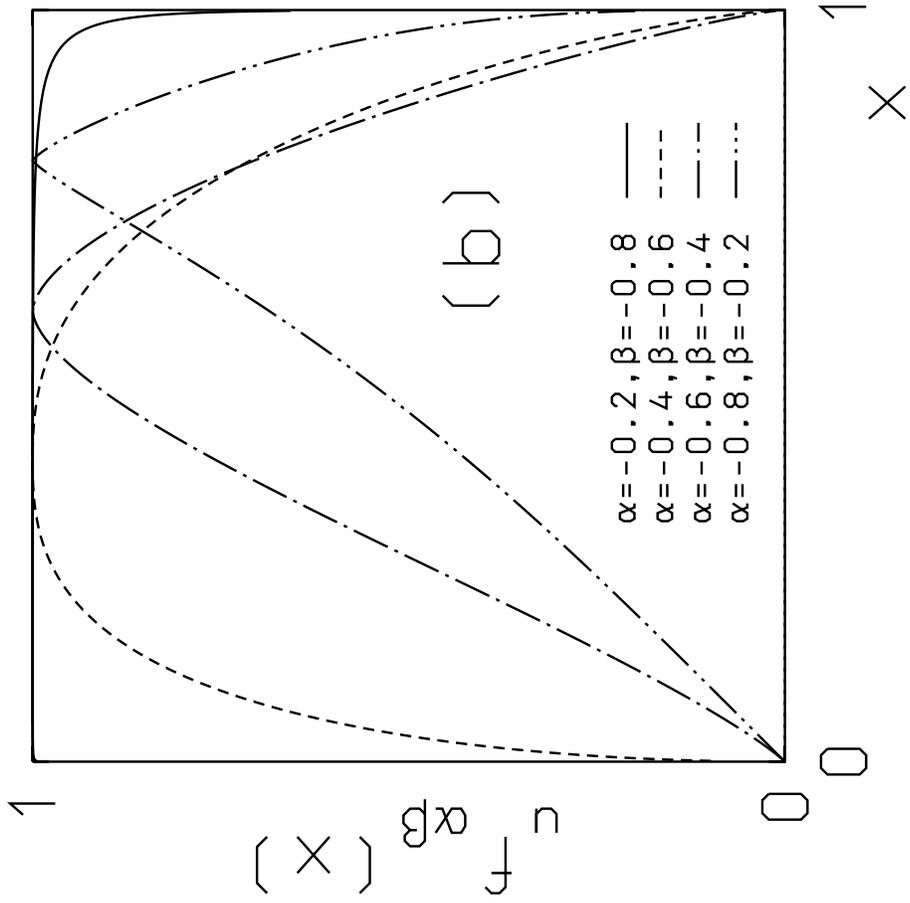
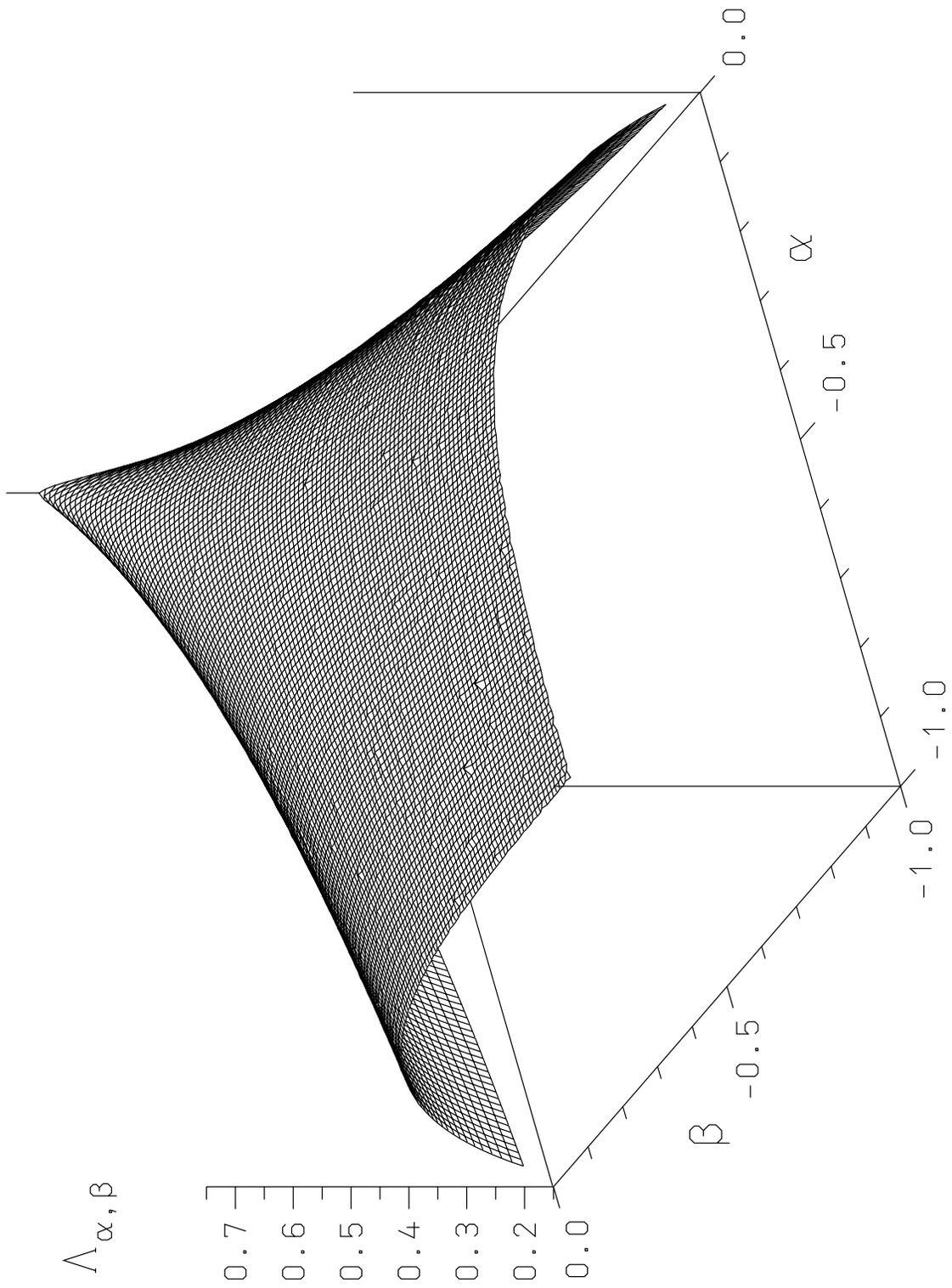
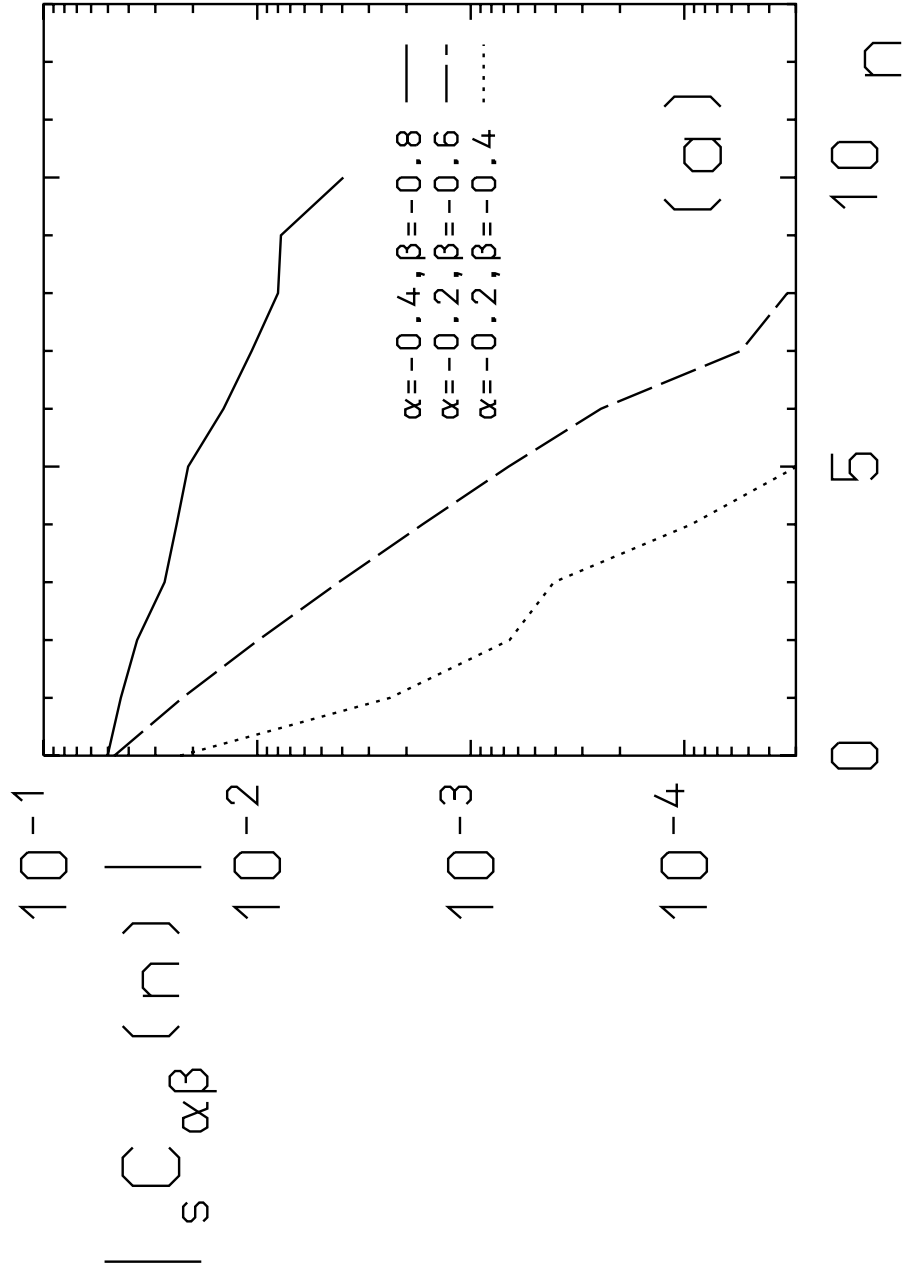
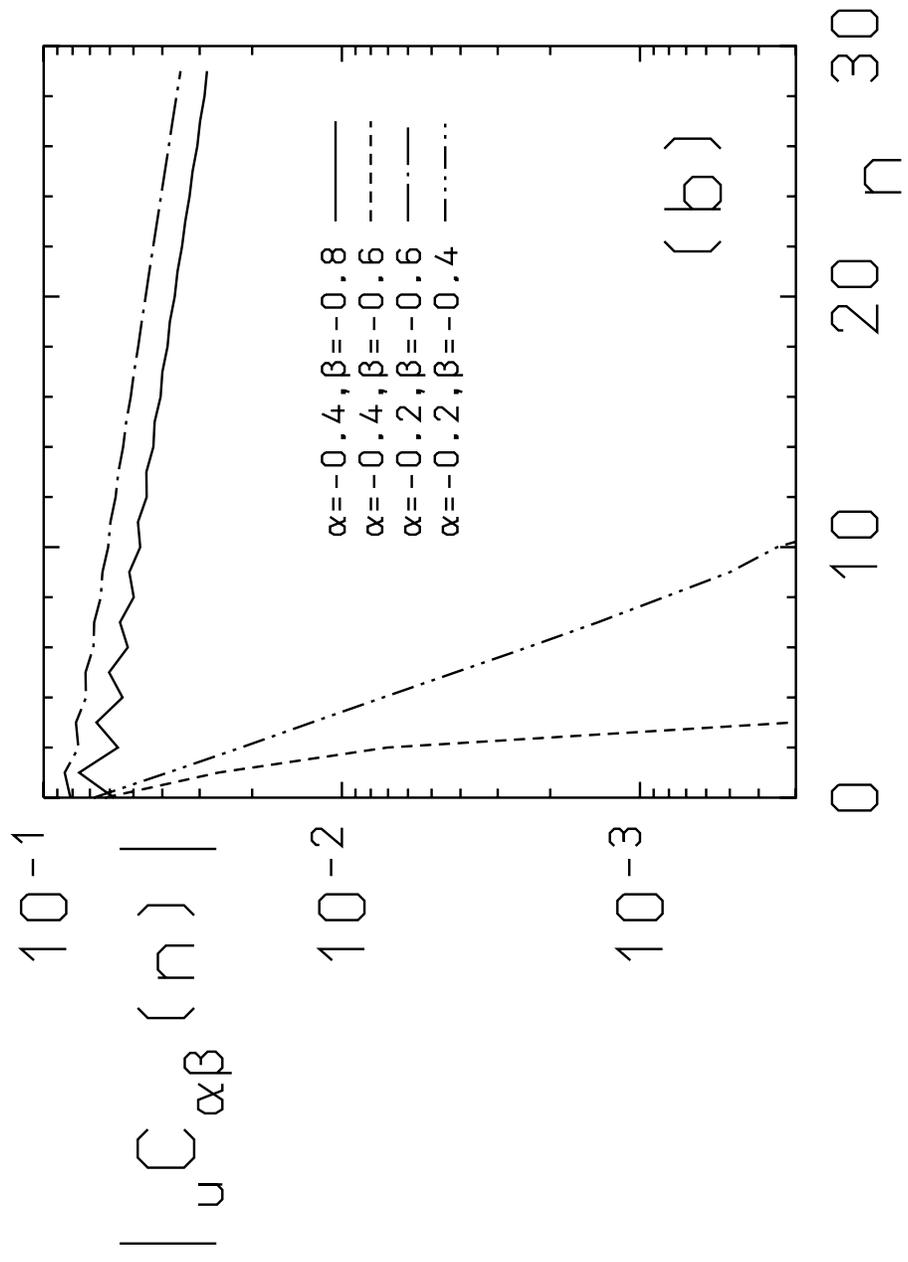
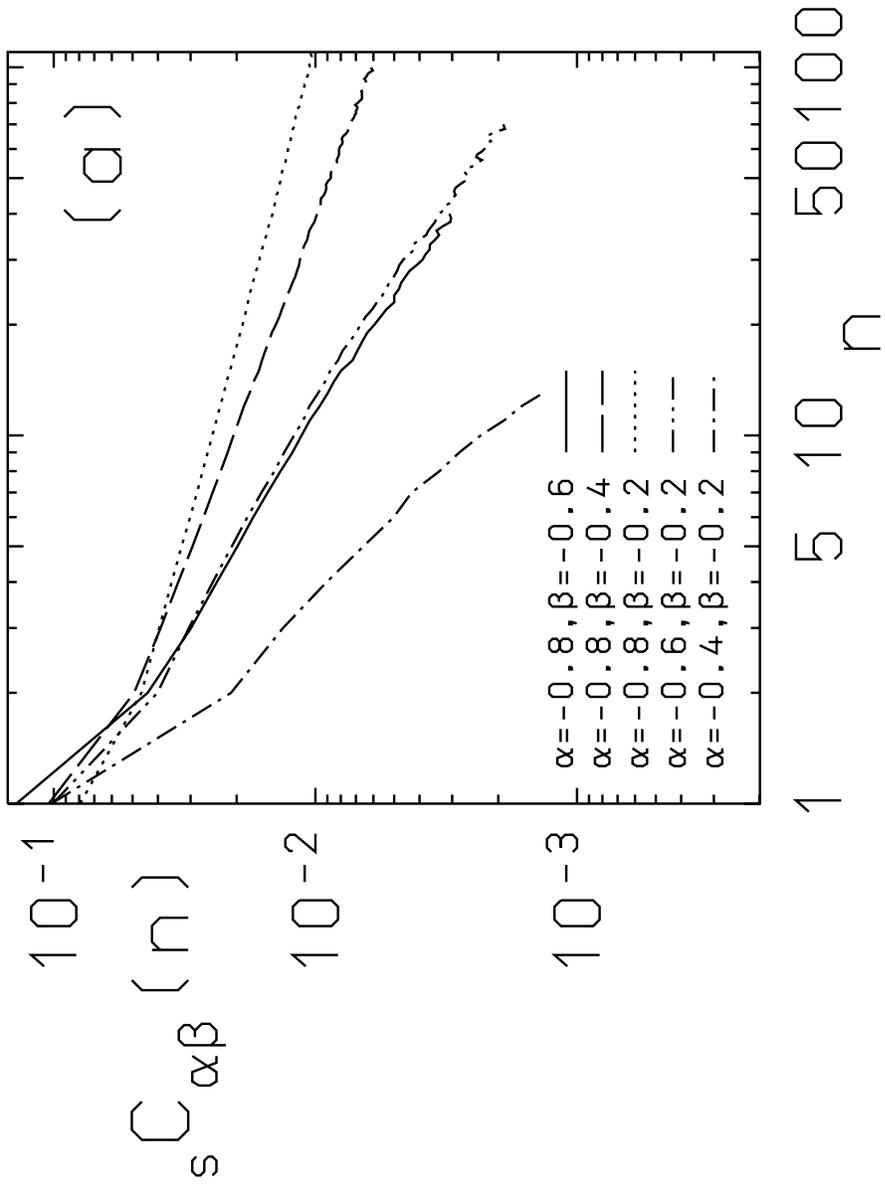
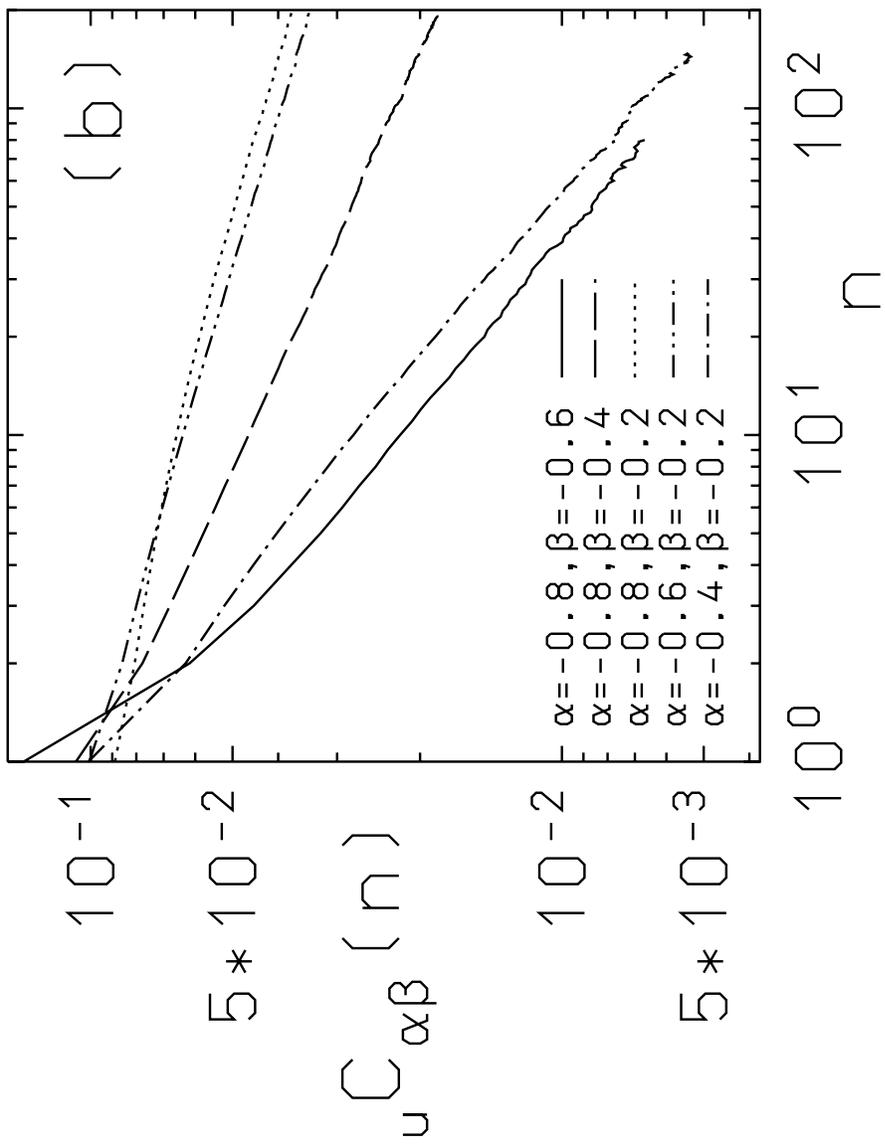
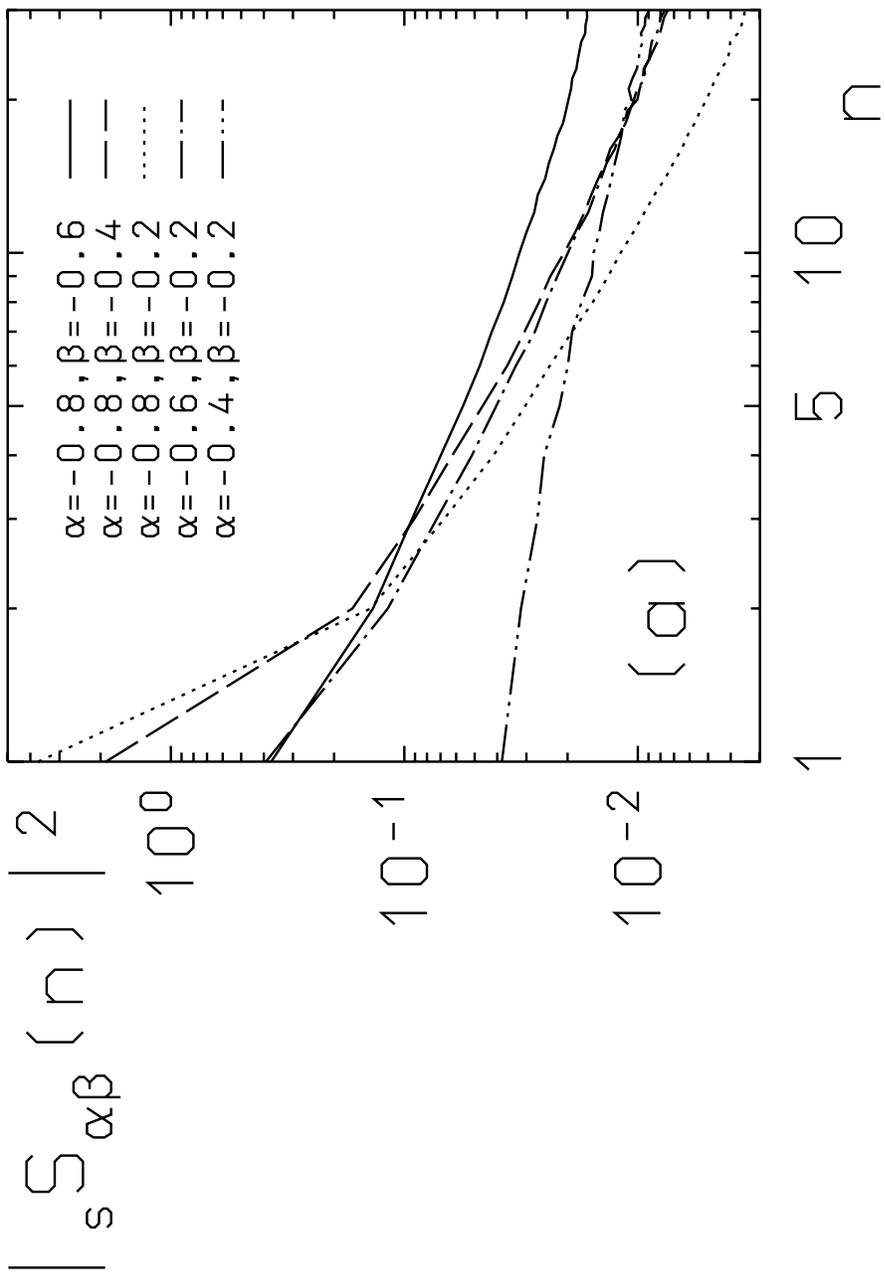
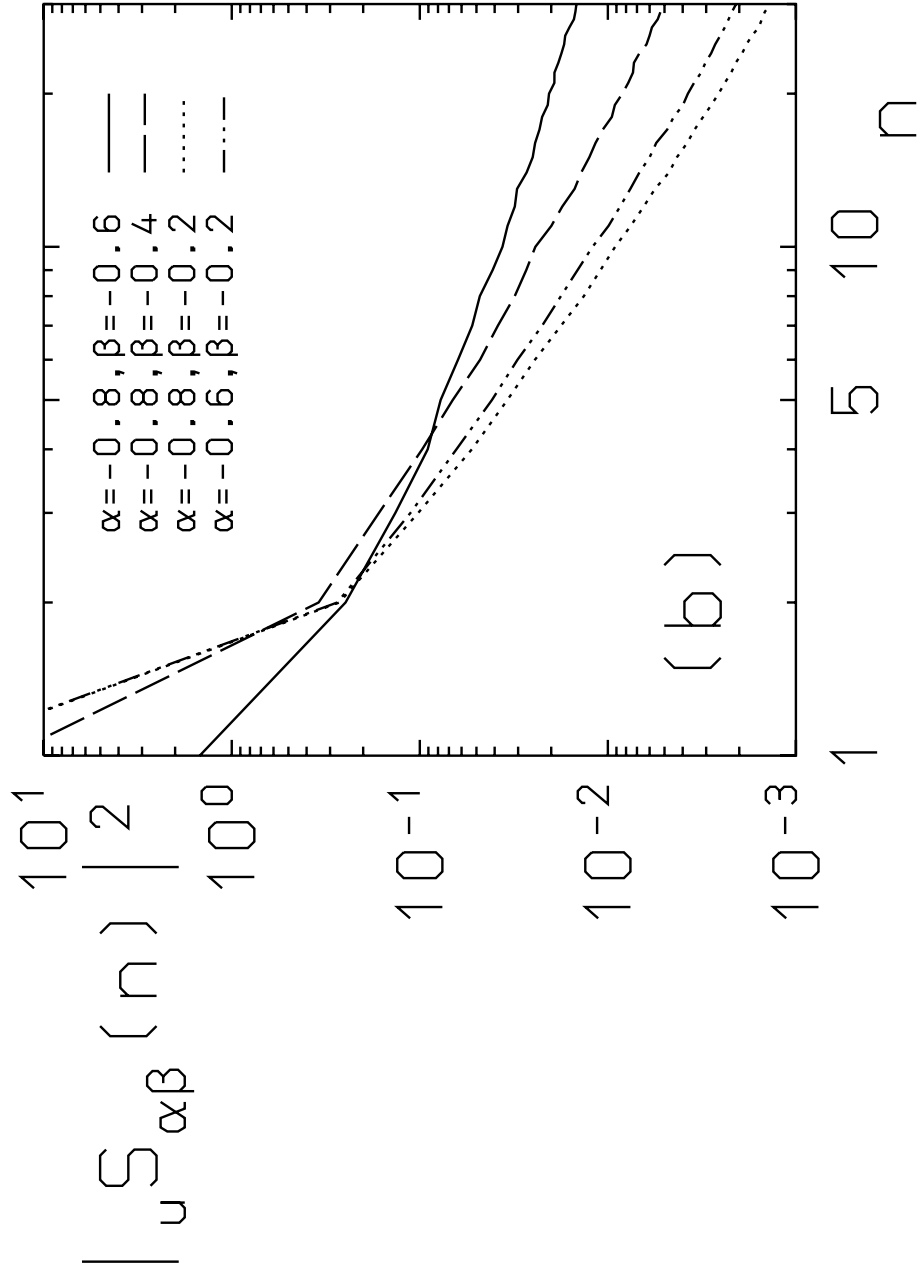
\end{document}